\documentstyle{mn}
\input epsf
\newcommand{\reff}{\mbox{$R_{\rm e}$}}

\newcommand{\msun}{\mbox{${\rm M}_{\odot}$}}
\newcommand{\ishape}{{\sc ishape}}
\newcommand{\daophot}{{\sc daophot}}
\newcommand{\daofind}{{\sc daofind}}
\newcommand{\tinytim}{{\sc TinyTim}}

\newcommand{\mfir}{\mbox{$m_{\mbox{\scriptsize FIR}}$}}
\newcommand{\bvz}{\mbox{$(\bv)_0$}}
\newcommand{\bv}{\mbox{$B\!-\!V$}}
\newcommand{\ub}{\mbox{$U\!-\!B$}}

\newcommand{\ubz}{\mbox{$(\ub)_0$}}

\newcommand{\tlu}{\mbox{$T_L(U)$}}
\newcommand{\tlv}{\mbox{$T_L(V)$}}
\newcommand{\apj}{ApJ}
\newcommand{\apjs}{ApJS}
\newcommand{\apjl}{ApJL}
\newcommand{\aap}{A\&A}
\newcommand{\aaps}{A\&AS}
\newcommand{\pasp}{PASP}
\newcommand{\aj}{AJ}
\newcommand{\araa}{ARA\&A}

\newcommand{\ssfr}{\mbox{$\Sigma_{\rm SFR}$}}
\begin{document}
\title{Young massive star clusters in M51}
\author[S. S. Larsen]{S{\o}ren S. Larsen \\
  UC Observatories / Lick Observatory, University of California \\
       Santa Cruz, CA 95064, USA \\
       email: soeren@ucolick.org}

\maketitle
\begin{abstract}
  A search for young massive star clusters (YMCs) in the nearby face-on
spiral galaxy M51 (NGC~5194) has been carried out using $UBV$ CCD
images from the prime focus camera on the Lick 3 meter Shane telescope.
The YMC population is found to be quite rich with a specific $U$-band 
luminosity $\tlu \sim 1.4$, consistent with the high current star formation 
rate of this galaxy. The brightest clusters 
have $M_V \sim -12.5$, far brighter than any young clusters currently
known in the Milky Way and even surpassing the luminosity of the R136 
cluster in the 30 Dor complex in the Large Magellanic Cloud. A few of the 
YMCs are examined on archive HST/WFPC2 images, confirming their cluster 
nature and providing estimates of their effective radii of $2-3$ pc. 
The number of YMCs in M51 is compatible with extrapolation of a power-law
luminosity function with exponent $\sim -2$ from a Milky Way-like 
population of open clusters.  Both the SFR and \tlu\ of M51 are similar 
to those of other cluster-rich spiral galaxies like NGC~1313 and M83.  
\end{abstract} 


\section{Introduction}

  M51 is probably among the most famous galaxies in the sky, instantly
recognizable by its nearby companion NGC~5195. Also known as
the ``Whirlpool Galaxy'', it was one of the first galaxies in which 
spiral structure was discovered by Lord Rosse, and it provides a 
spectacular text-book example of a grand-design Sbc type spiral 
seen nearly face-on.  It has a very high optical surface brightness
\cite{oka76} and is also a strong emitter of far-infrared radiation, 
indicating strong star formation activity in the spiral arms 
\cite{smi82,hip96}. 
Deep images reveal faint outlying material embedding both M51 and NGC~5195 
\cite{bur78}, providing clear evidence that the two galaxies are physically 
interacting.  These characteristics, along with the relatively 
small distance ($8.4 \pm 0.6$ Mpc, Feldmeier et al.~1997) and the location 
far from the galactic plane ($b = 68^{\circ}$) make M51 an attractive target 
for studying its population of {\it Young Massive Star Clusters} (YMCs). 

  YMCs are abundant in interacting and merger galaxies like e.g. the
``Antennae'' NGC~4038/4039 (a list of galaxies with known YMC populations
is given in Larsen (1999b)).  However, they are also seen in more normal 
galaxies like e.g. the LMC \cite{van91} and M33 \cite{chr88}.
In a recent study of 21 nearby, 
non-interacting spiral galaxies, Larsen \& Richtler \shortcite{lar00} 
(LR2000) found a strong 
correlation between the specific $U$-band luminosity \tlu\ of YMCs 
in a galaxy and the area-normalized star formation rate \ssfr .
In this respect, M51 provides a highly interesting intermediate case of a
clearly interacting galaxy that has still retained the characteristics of 
a normal spiral.

  This paper reports the results of a study of YMCs in M51.
In Sect.~\ref{sec:selection}, YMCs are identified and photometry is 
obtained from ground-based $UBV$ CCD imaging.  Archive HST/WFPC2 images 
are then used to examine a few clusters in detail, including measurements 
of their sizes (Sect.~\ref{sec:hst}). Next, the specific $U$-band luminosity 
of the M51 cluster system is derived and compared with other galaxies 
(Sect.~\ref{sec:glob}).  In Sect.~\ref{sec:disc} the YMC population 
in M51 is compared with young clusters in the Milky Way and the LMC,
along with some considerations on formation of YMCs.  Finally, conclusions 
are in Sect.~\ref{sec:conc}.

\section{Data}

\subsection{Observations and initial reductions}

  CCD images in the $UBV$ passbands were obtained on Mar 13 -- 14, 2000
with the prime focus camera (PFCAM) on the Lick 3 meter Shane telescope at
Mount Hamilton, California.  The total integration times were 3600, 1200
and 900 sec. in $U$, $B$ and $V$, split into 3 individual exposures in
each filter.  Typical seeing values ranged between 
1\farcs0 and 1\farcs5, but the image quality was
degraded by problems with the alignment of the mirror cell after recent
re-aluminization and oscillations due to wind, making stellar images 
appear somewhat elongated. The image scale was 0\farcs296 /pixel 
and the total field of view was $10^\prime \times 10^\prime$, sufficient 
to fully cover both M51 itself and the companion NGC~5195.

  To reduce read-out time, the two halves of the CCD were read out in
parallel through two amplifiers.  During the subsequent reductions, the 
difference between the bias levels in the two parts of the CCD image 
was found to change by up to $\sim 40$ ADU from one exposure to another.
The difference was eliminated by adding a constant number to all pixel
values in one half of the image. A more serious concern was that the
flat-field varied significantly, most likely because of scattered light.  
Dividing two skyflats in the same filter with each other, the ratio was 
found to vary by up to $\sim20\%$ from the centre to the corner of the 
chip, with the most severe problems in the southern $\sim300$ pixels 
(about $1\farcm5$) of the field. Within the central $1000 \times 1000$ 
pixels the variations were less dramatic, but still quite significant at 
about the 5\% level. The effect was clearly visible in the final 
calibrated images as large-scale gradients in the background. 
However, the pattern appeared to be more or less the same in all passbands 
so the flatfielding errors may, to a certain extent, cancel out for
colour indices.

\subsection{Photometric calibration}

\begin{table}
\caption{\label{tab:hstdat} 
  HST archive images of M51 used for photometric
 calibration of ground-based photometry.}
\begin{tabular}{cccc} 
  PID & PI & Filter & Exposure times \\ \hline
5777 & Kirschner & F555W & 600 s \\
5777 & Kirschner & F439W & $2 \times 700$ s \\
5652 & Kirschner & F336W & $3 \times 400$ s \\ \hline
\end{tabular}
\end{table}

  Standard fields from Landolt \shortcite{lan92} were observed for photometric 
calibration. The transformation equations were assumed to be of the form
\begin{eqnarray*}
  V   & = & v + c_v \times (b-v) + z_v  \\
  \bv & = & c_{(b-v)} \times (b-v) + z_{(b-v)} \\
  \ub & = & c_{(u-b)} \times (u-b) + z_{(u-b)} 
\end{eqnarray*}
where capital letters denote standard magnitudes and small letters indicate
instrumental magnitudes, corrected for atmospheric extinction.  However, 
because of non-photometric conditions 
and the flat-field problems it was necessary to correct the zero-points of 
the $UBV$ photometry using HST archive images. $UBV$ magnitudes were obtained 
for sources in the HST images listed in Table~\ref{tab:hstdat}, following the 
standard procedure described in Holtzman et al.~\shortcite{hol95}. Objects 
bright enough to be 
used for calibration of the ground-based images were easily measured through 
the Holtzman et al.  reference aperture of $r=5$ pixels, avoiding problematic 
aperture corrections.  The zero-points ($z_v$, $z_{(b-v)}$ and $z_{(u-b)})$ 
of the ground-based magnitudes were then adjusted to fit the HST system, while 
the scaling constants ($c_v$, $c_{(b-v)}$ and $c_{(u-b)}$) derived from the 
Landolt standard fields were preserved. The overlap between the HST datasets 
used for the $BV$ and $U$ calibrations amounted to less than the area of one 
WF camera chip, making it difficult to find a suitable number of calibration
objects. The \ub\ calibration is thus somewhat more uncertain than the
\bv\ and $V$-band calibrations and it is estimated that $z_{(u-b)}$ is
good to $\pm 0.15$ mag while $z_{(b-v)}$ is probably accurate to
better than $\pm 0.05$ mag. Although significantly worse than what could
have been obtained under optimal photometric conditions, the calibrations
adopted here are nevertheless accurate enough for the conclusions drawn 
in this paper.

\section{Selection of cluster candidates}
\label{sec:selection}

  In order to facilitate a comparison with the results of LR2000,
the same data reduction and analysis procedures were followed as closely
as possible.  Objects were detected using the \daofind\ task in
\daophot\ \cite{ste87} on background-subtracted $V$ and $B$ band
images and matching the two object lists.  Aperture photometry was then
carried out as described in Larsen \shortcite{lar99a}, using an aperture 
radius of $r=5$ pixels.  Finally, a correction for interstellar absorption 
of $A_B = 0.150$ mag was applied \cite{sch98}.

\begin{figure}
\epsfxsize=84mm
\epsfbox[82 371 549 718]{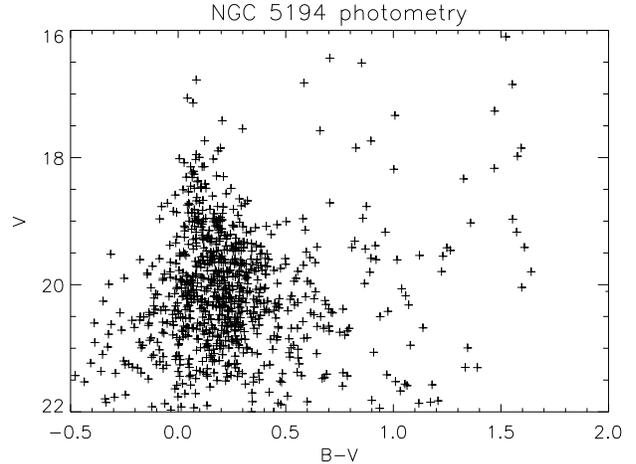}
\caption{\label{fig:bv_v}
  (\bv , $V$) colour-magnitude diagram for objects in the M51 field. Objects
with $\bv > 0.5$ are mostly foreground stars whereas contamination blueward
of $\bv=0.5$ is expected to be small.}
\end{figure}

The resulting ($\bv, V$) colour-magnitude diagram is shown in 
Fig.~\ref{fig:bv_v}. An initial list of cluster candidates was 
then obtained by selecting all objects with 
\begin{itemize}
  \item $\bv < 0.45$
  \item $ M_V < \left\{ \begin{array}{ll}
                      -9.5 & \mbox{for } \ub < -0.4 \\
                      -8.5 & \mbox{for } \ub \ge -0.4 \\
                 \end{array}
             \right.  $
\end{itemize}
  These criteria ensure minimal contamination from foreground stars, the
majority of which will be redder than the \bv\ cut, and by individual
luminous stars in M51 itself because of the $M_V$ limit.  However, 
many of the objects identified in this way were still not likely cluster 
candidates, but high-surface brightness parts of spiral arms, HII regions 
or just artefacts found by {\sc daofind} where no real object existed. 
A visual inspection of the images, looking for relatively isolated, 
point-like sources, was found to be the only way to isolate good cluster 
candidates from the multitude of other objects in the frames.

\begin{figure}
\epsfxsize=84mm
\epsfbox[82 371 547 717]{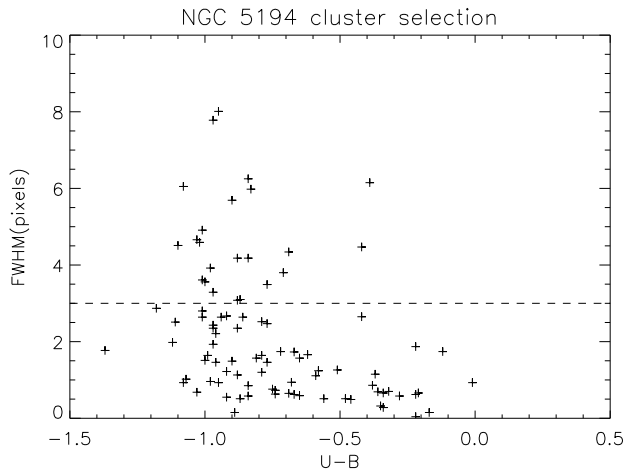}
\caption{\label{fig:ub_sz}
  The intrinsic FWHM vs. \ub\ colour for cluster candidates in M51 that
passed the initial visual inspection.  The horizontal dashed line 
indicates the size cut applied to further constrain the cluster 
sample.}
\end{figure}

  LR2000 further used $H\alpha$ images to reject the youngest objects, 
still embedded in HII regions and often surrounded by a fuzz of nebular 
emission and star forming regions, but no $H\alpha$ filter was available at 
PFCAM during this observing run. Instead, such objects were rejected
based on their angular sizes.  Object sizes were measured using 
the \ishape\ algorithm \cite{lar99a}, convolving the point spread function
(PSF) of the $V-$band image with King profiles
to match the observed object profiles. A concentration parameter of
30 was assumed for the King profiles.  Fig.~\ref{fig:ub_sz} shows the 
intrinsic object sizes derived by \ishape\ as a function of \ub\ colour for 
objects that passed the initial visual inspection. Clearly, most of the 
more extended sources are found among the bluest objects, while most 
objects with $\ub \ga -0.6$ have FWHM $<$ 2 pixels. 
Note that the typical seeing in our images corresponds to FWHM $\sim 5$ 
pixels.  
A size cut at FWHM=3 pixels was 
finally applied to exclude the fuzziest objects, likely to be OB 
associations or HII regions rather than real star clusters.

  Although the cluster colours have been corrected for Galactic foreground
extinction, some residual reddening may still be present from M51 itself.
The question of the optical thickness of face-on spiral galaxies is still 
debated, and for any particular object the absorption will obviously depend 
on the local structure of the surrounding interstellar medium.  However,
a recent study \cite{xil99} concluded that face-on spiral galaxies are
essentially transparent with an optical depth of less than one in all 
optical bands. M51 being obviously face-on (inclination angle
$\sim 20^\circ$, Tully 1974), interstellar extinction problems are thus 
expected to be small.

\begin{figure}
\epsfxsize=84mm
\epsfbox[79 376 542 714]{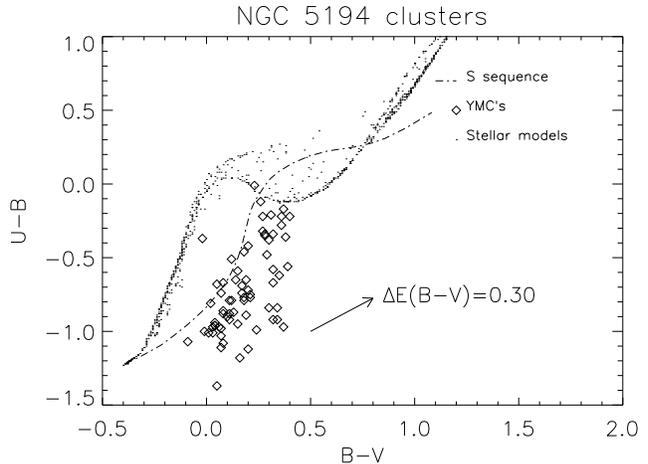}
\caption{\label{fig:bv_ub}
  (\bv , \ub) two-colour diagram for the final sample of cluster 
  candidates in M51}
\end{figure}

  The final list consists of 69 cluster candidates (Table~\ref{tab:clusters}).
A ($\bv, \ub$) diagram for these is shown in 
Fig.~\ref{fig:bv_ub}, along with stellar models by Bertelli et 
al.~\shortcite{ber94} and the Girardi et al.~\shortcite{gir95} 
``S''--sequence which marks the average colours of 
LMC clusters. The arrow indicates the effect of a reddening of
0.3 in \bv, equivalent to 0.93 magnitudes of $V$--band absorption for a 
standard reddening law.  The M51 clusters tend to have somewhat bluer \ub\ 
colours than expected from the S-sequence, but a shift of about 0.1 in 
\ub\ would make them match the S-sequence colours nicely. Such a shift 
would be consistent with the estimated uncertainties. The S-sequence is
essentially an age sequence and Girardi et al.~\shortcite{gir95} published 
a calibration of age as a function of position in the ($\bv,\ub$) diagram. 
However, because of the uncertainties in the photometric calibration 
and especially in the $U$ band, a detailed discussion of ages for the M51 
clusters will not be given in this paper, but note that the clusters do 
scatter along the S-sequence, indicating a significant age spread.  Because 
of the fading with age, the sample becomes increasingly incomplete for 
higher ages.  The upper age limit, due to the \bv\ cut, is about 500 Myr.

\begin{figure}
\epsfxsize=84mm
\epsfbox{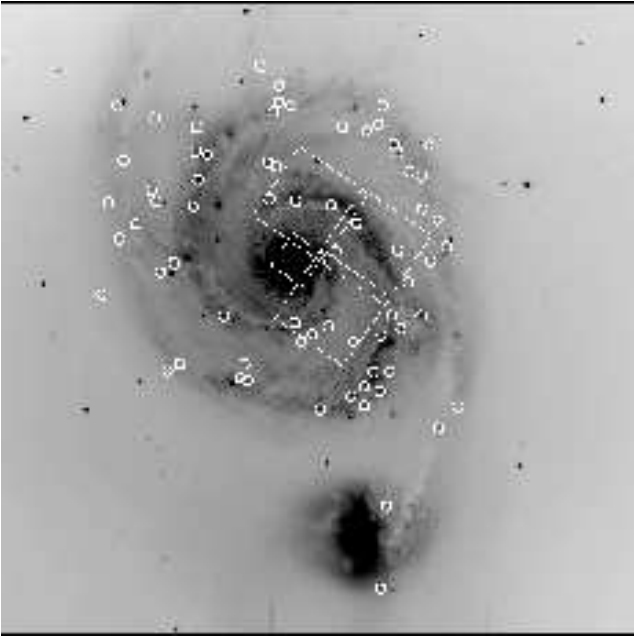}
\caption{\label{fig:m51v}
  A $V-$band image of M51 with the cluster positions
indicated. South is up.  The large-scale variations 
in the background are due to poor flatfielding, presumably because of 
scattered light in the skyflats.  The F439W / F555W HST pointing 
(PID 5777) is also shown.
}
\end{figure}

  Fig.~\ref{fig:m51v} shows a $V$-band image with the cluster positions
indicated. Most of the cluster candidates are located in or near the spiral
arms of M51 itself. This is to a large extent a selection effect, due to the 
fact that the youngest clusters are more luminous (for any given mass) and 
therefore easier to detect.  However, note that also the companion galaxy 
contains a couple of massive clusters. The HST F439W + F555W pointing is
also shown in Fig.~\ref{fig:m51v}.

\section{HST data}
\label{sec:hst}

\begin{table*}
\caption{\label{tab:hstclust} Comparison of ground-based and
  HST-based data for 7 clusters in M51. Cols. 2 and 3 list ground-based
  $V$ and \bv\ magnitudes, Cols. 4 and 5 give the corresponding values
  measured on HST WFPC/2 images and the differences
  $\Delta V = V(\mbox{ground}) - V(\mbox{HST})$ and
  $\Delta \bv = \bv (\mbox{ground}) - \bv (\mbox{HST})$ are in Cols. 6 and 7.
  The colours given here have not been corrected for reddening.
  Effective cluster radii measured on the HST images are given in the last
  column.
}
\begin{tabular}{rccccrrr} 
  ID & $V$ (ground) & \bv\ (ground) & $V$ (HST) & \bv\ (HST) & $\Delta V$ &
  $\Delta \bv$ & \reff\ (pc) \\ \hline
  403 & 19.58 & 0.342 & 19.80 &    0.29 & $-0.22$ & $0.05$  & 2.7 \\
  477 & 18.41 & 0.210 & 18.46 &    0.33 & $-0.05$ & $-0.12$ & 0.9 \\
  479 & 18.11 & 0.098 & 18.83 &    0.06 & $-0.72$ & $0.04$  & 3.8 \\
  534 & 19.48 & 0.177 & 20.24 & $-$0.04 & $-0.76$ & $0.22$   & 2.3 \\
  589 & 19.40 & 0.070 & 19.51 &    0.08 & $-0.11$ & $-0.01$  & 2.1 \\
  597 & 19.90 & 0.177 & 19.72 &    0.52 & $ 0.18$ & $-0.34$  & -   \\
  627 & 20.13 & 0.237 & 20.53 &    0.51 & $-0.40$ & $-0.27$  & 0.0 \\
  693 & 19.42 & 0.148 & 19.66 &    0.11 & $-0.24$ & $0.04$  & 2.7 \\
  713 & 18.33 & 0.071 & 19.22 & $-$0.11 & $-0.89$ & $0.18$  & 0.8 \\
  912 & 19.78 & 0.317 & 20.34 &    0.16 & $-0.56$ & $0.16$  & 2.2 \\ \hline
\end{tabular}
\end{table*}

\begin{figure}
\epsfxsize=85mm
\epsfbox{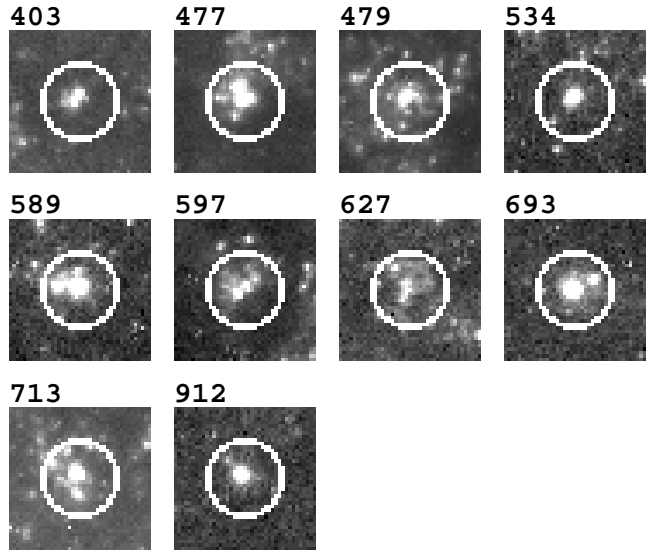}
\caption{\label{fig:hstclust}
  Clusters identified on HST images. The circles are 2 arseconds in diameter.
}
\end{figure}

  In addition to being useful for calibrating the photometry, the 
superior angular resolution of the HST data provides a welcome way of
examining the structure of the cluster candidates in much greater detail
than what is possible from the ground. Fig.~\ref{fig:hstclust} shows
WFPC/2 close-ups of 10 clusters, marked by asterisks ($*$) in 
Table~\ref{tab:clusters}.  The images in Fig.~\ref{fig:hstclust} are from 
two combined F439W exposures, allowing for cosmic ray rejection.

  First, the HST images clearly confirm the cluster nature of most of 
the objects. Exceptions may be objects \#627 -- a quite faint source
located near the nucleus of M51, and \#597 and \#403 which may consist
of two or more closely separated individual stars.  The remaining seven 
objects are far too bright to be individual stars and appear well resolved.

  Most of the clusters are evidently not isolated objects, but are surrounded 
by individual luminous stars and other clusters. This is, however, to be 
expected because most of these clusters have very blue \ub\ colours, 
indicating low ages.  A \ub\ colour of $-0.8$, typical for the clusters in
Fig.~\ref{fig:hstclust}, corresponds to an age of about 12 Myr 
\cite{gir95}, too short for a young star cluster to escape from the 
environment where it was born. The two reddest clusters (\#693 and \#912) 
with $\ub = -0.59$ and $\ub = -0.58$ already appear somewhat more isolated 
than the other clusters in the Fig.~\ref{fig:hstclust}, consistent with 
the higher ages inferred from their \ub\ colours ($\sim 30$ Myr). 


\subsection{HST versus ground-based photometry}

  Table~\ref{tab:hstclust} lists HST-based photometry and sizes (see 
Sect.~\ref{sec:hstsz})  for the 10 objects in Fig.~\ref{fig:hstclust}.  
For comparison, ground-based photometry is also listed. The photometry
in Table~\ref{tab:hstclust} has not been reddening corrected.  

  Magnitudes measured on the ground-based images are generally brighter 
than those based on HST images.  This is not surprising, considering that 
the HST magnitudes were measured through a 0\farcs5 aperture 
which clearly does not include all the light surrounding the clusters that 
will enter into a ground-based aperture.  Generally, the agreement on 
\bv\ colours is better than on $V$ magnitudes, confirming that colours can 
be more accurately measured than individual magnitudes for objects in 
crowded fields. The difference between ground-based and HST-based \bv\ 
colours clearly increases for the fainter objects, with the largest 
discrepancy for the likely ``non-clusters'' (\#597 and \#627) pointed 
out above.

  It is also worth noting that the HST pointings were centred on the nucleus 
of M51, providing a ``worst-case'' comparison between HST and ground-based 
data. Further out in the disk of M51 the surface brightness drops and the 
cluster sample becomes less biased towards the brightest (and thus youngest) 
objects where crowding problems are most significant.

\subsection{Cluster sizes} 
\label{sec:hstsz}

  Sizes were measured on the WFPC/2 F439W images for the clusters in
Fig.~\ref{fig:hstclust} with the \ishape\ algorithm \cite{lar99a}.
At the distance of M51, one WF pixel corresponds to a linear 
scale of 4 pc, so measuring cluster sizes of the order of a few pc
is not trivial. The input PSF
for \ishape\ was generated empirically from point sources in the HST 
images, using the {\sc PSF} task in \daophot . This was preferred over
generating the PSFs with the \tinytim\ PSF simulater \cite{kri97} which
does not include the so-called ``diffusion kernel'' for subsampled PSFs, 
and would thus have caused \ishape\ to derive too large cluster sizes.

\begin{figure}
\epsfxsize=84mm
\epsfbox[84 368 540 716]{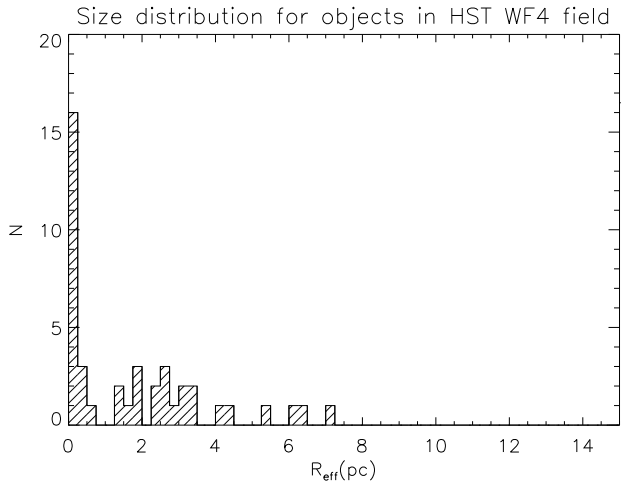}
\caption{\label{fig:szhist}Histogram of the size distribution
 for objects in the WF4 chip of the combined F439W image.}
\end{figure}

  The clusters were modelled as King profiles with concentration parameter
$c=30$, resulting in the half-light radii (\reff ) listed in the last column 
of Table~\ref{tab:hstclust}.  The cluster sizes measured here are consistent 
with sizes for YMCs in merger galaxies like the Antennae \cite{whi99}, 
with $\reff \sim 2 - 3$ pc and do also match the typical size of 
{\it globular clusters} in the Milky Way and other galaxies quite well 
\cite{har96}.  It has been estimated from simulations that
\ishape\ is able to measure sizes down to $\frac{1}{10}$ times the size 
of the PSF when sufficient signal is present \cite{lar99a}, corresponding 
to $\sim 0.5$ pc in the current case. To illustrate the reliability of
the cluster sizes, Fig.~\ref{fig:szhist} shows a 
histogram of the size distribution for all objects brighter than $V=22$ 
in the WF4 chip. The histogram shows a narrow peak at $\reff = 0$,
corresponding to unresolved point sources, while objects with sizes 
characteristic of the YMCs appear well separated from the $\reff=0$ peak.
The YMCs are thus well resolved and their sizes are well determined.

\section{Global properties of the M51 cluster system}
\label{sec:glob}


\subsection{Specific luminosity}

  In studies of old globular cluster systems (GCSs) it has become customary
to characterize the richness of a GCS by its {\it specific frequency},
originally defined by Harris \& van den Bergh \shortcite{har81} as the 
number of globular clusters
($N_{\mbox{\scriptsize GC}}$) per host galaxy luminosity ($M_V$):
$S_N = N_{\mbox{\scriptsize GC}} \times 10^{0.4 \times (M_V + 15)}$.
However, a potentially better measure is the specific {\it luminosity}
\cite{har91}, which is the ratio of the total luminosity of the cluster
system to that of the host galaxy.  Specific luminosities are much less 
sensitive to incompleteness effects at the lower end of the cluster 
luminosity function because the fainter clusters obviously contribute less 
by light than they do by number.  For YMC systems this advantage becomes 
even greater than for old GCSs because the faint wing of the cluster 
luminosity function is usually unknown, so a definition of an equivalent 
specific frequency for YMCs has to rely on artificial and somewhat 
arbitrary magnitude limits.

  LR2000 defined the specific $U$-band luminosity for YMCs as follows:
\begin{equation}
 T_L(U) = 100 \cdot \frac{L_{\rm Clusters}(U)}{L_{\rm Galaxy}(U)},
  \label{eq:tu}
\end{equation}
where $L_{\rm Clusters}(U)$ and $L_{\rm Galaxy}(U)$ are the total $U$-band
luminosities of the cluster system and of the host galaxy, respectively. 
The $U$ band was chosen to sample the {\it young} cluster and stellar 
populations in a galaxy as cleanly as possible. 


\begin{figure}
\epsfxsize=84mm
\epsfbox[68 368 547 717]{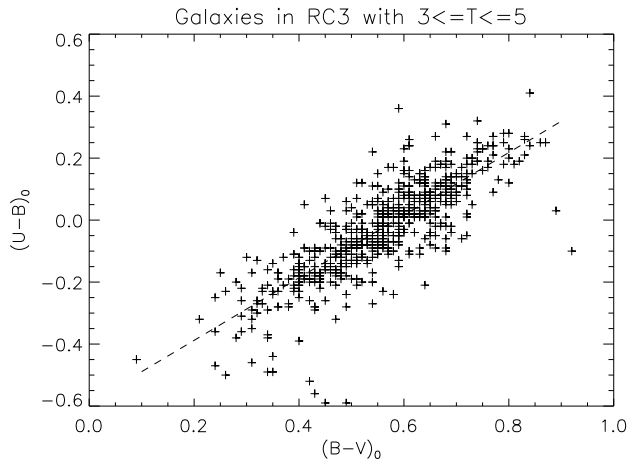}
\caption{\label{fig:bv_ub_rc3}
  \ub\ vs. \bv\ for galaxies with $T$-type in the range 3 to 5 (from RC3).
  The dashed line is a fit to the data points.
  }
\end{figure}

  To calculate \tlu\ for M51, $L_{\rm Clusters}(U)$ was obtained 
simply by adding up the $U$-band light from all clusters in 
Table~\ref{tab:clusters}.  Unfortunately, no integrated $U$-band 
photometry is available for M51 itself so its $U$-band magnitude 
was {\it estimated} from the $BV$ photometry given in the 
RC3 catalogue \cite{vau91}. Fig.~\ref{fig:bv_ub_rc3} is a plot of 
reddening-corrected
\ubz\ vs. \bvz\ for all galaxies in RC3 with $T$ types between 3 and 5, 
M51 itself being classified as Sbc ($T=4$). The data in 
Fig.~\ref{fig:bv_ub_rc3} can be fitted by a straight line, yielding
\begin{equation}
  \ubz = 1.01\times\bvz - 0.59
  \label{eq:bv_ub}
\end{equation}
with a scatter of $\sigma(\ub) = 0.09$ around the fit. The integrated
\bvz\ colour for M51 is given as 0.53 in RC3, so from 
Eq.~(\ref{eq:bv_ub}) we get $\ubz = -0.05$. This should be accurate to
0.1 magnitudes, translating into a 10\% uncertainty on the
derived \tlu\ value.
These \bvz\ and
\ubz\ colours also compare well with integrated photometry for NGC~5236 (M83), 
a galaxy which is in many respects quite similar to M51. For NGC~5236, 
RC3 gives integrated \bvz\ and \ubz\ colours of 0.61 and $-0.01$, respectively,
i.e. M51 is slightly bluer in both \bv\ and \ub .

  The specific $U$-band luminosity of the M51 YMCs can now be
obtained:
\begin{equation}
  \tlu = 1.36 \pm 0.28
\end{equation}
  Similarly, the $V$-band specific luminosity is found to be
$\tlv = 0.41\pm0.07$. 
The errors quoted here are the formal errors due to sample statistics. 
The real uncertainties are probably higher, but are 
hard to estimate.  For example, lowering the size cut to FWHM=2 pixels 
instead of 3 pixels would change $\tlu$ to $0.95\pm0.24$.  On the
other hand, the M51 photometry does not go quite as deep as most of
of the galaxies in the sample of LR2000, -- primarily because M51 is 
further away.  Thus \tlu\ and \tlv\ are
probably both somewhat underestimated. \tlv\ may be more affected 
because older, redder clusters carry more weight than for \tlu , while
at the same time being fainter and therefore more incomplete.

\subsection{The \tlu\ vs. \ssfr\ correlation}

  LR2000 found a very good correlation between \tlu\ and the
area-normalized star formation rate \ssfr\ for the 21 galaxies in their
sample and an additional 10 galaxies (mostly starbursts and merger 
galaxies) for which data were collected from the literature. How does
M51 fit into this relation?

  To derive \ssfr\ for M51, we use the relation from LR2000:
\begin{equation}
  \ssfr (\msun \, {\rm yr}^{-1} \, {\rm kpc}^{-2})
     = 144000 \times 10^{-0.4 \, \mfir \, - \, 2 \, \log D_0},
     \label{eq:ssfr}
\end{equation}
based on the calibration in Buat \& Xu \shortcite{bua96}.  \mfir\ is the 
RC3 FIR magnitude (based on IRAS $60\mu$ and $100\mu$ flux densities) and 
$\log D_0$ is the logarithm of the optical galaxy diameter 
in tenths of arcminutes.  Note that both \ssfr\ and \tlu\ are distance 
independent. In some cases the RC3 ``effective aperture''
($\log A_e$), containing half the $B$-band light, may provide a more
reasonable measure of the area to which the SFR should be normalized,
but this is not catalogued for all galaxies in RC3 and thus $\log D_0$
was chosen for homogeneity.
  
  For M51, $\mfir = 7.86$ and $\log D_0 = 2.05$. Inserting these
numbers in (\ref{eq:ssfr}), the star formation rate surface density is
then
\begin{equation}
  \ssfr = 8.21 \times 10^{-3} \msun \, {\rm yr}^{-1} \, {\rm kpc}^{-2}
\end{equation}
  This number is roughly a factor of two lower than the value given in
Kennicutt \shortcite{ken98b}, derived from the H$\alpha$ luminosity. 
Considering the intrinsic uncertainties in both methods, the agreement is 
quite satisfactory.

\begin{figure}
\epsfxsize=84mm
\epsfbox[80 372 547 715]{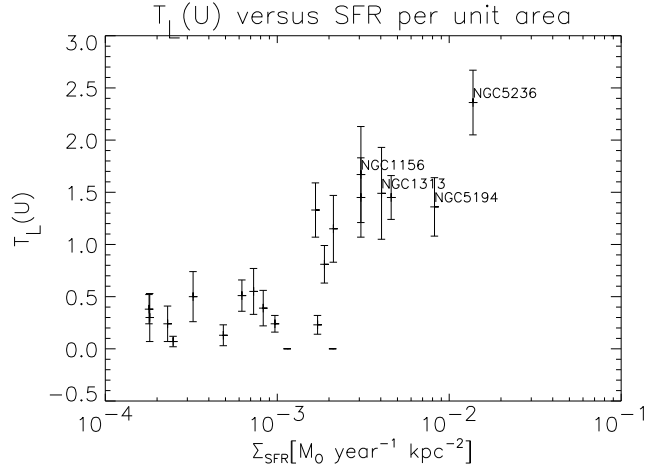}
\caption{\label{fig:sfr_tlu}
  Specific $U$-band luminosity as a function of star formation rate
surface density for galaxies in the sample of Larsen \& Richtler (2000)
and M51 (NGC~5194).
}
\end{figure}

  The \tlu\ -- \ssfr\ plot in LR2000 can now be updated with one more 
data point for M51, as shown in Fig.~\ref{fig:sfr_tlu}. M51 (labelled
by its NGC number, 5194) fits quite nicely into the existing relation, 
with a \tlu\ comparable
to other high \ssfr\ galaxies like NGC~1156 and NGC~1313. It may fall
somewhat below the \tlu\ value expected for its \ssfr , but the error bars 
in the horizontal direction of Fig.~\ref{fig:sfr_tlu} are 
probably substantial, with an uncertainty of at least 50\% on the calibration
of star formation rate as a function of FIR luminosity alone \cite{bua96}. 


\section{Discussion}
\label{sec:disc}

\subsection{Comparison of the luminosity functions of YMCs in M51 and 
  open clusters in the Milky Way}

M51 contains a large number of very luminous star clusters, including 
40 clusters with $M_V<-10$ and 4 clusters with $M_V < -12$ 
(Table~\ref{tab:clusters}).  As indicated by Table~\ref{tab:hstclust}, 
the ground-based $V$ magnitudes are systematically brighter
than the HST photometry, with an average difference of $\Delta V = -0.38$
for the objects listed in Table~\ref{tab:hstclust}. Applying a correction
of 0.38 mag to all $V$ magnitudes, M51 still contains 30 clusters with
$M_V<-10$ and $3$ with $V<-12$.  For comparison, the brightest 
known young clusters in the Milky Way have $M_V\sim-10$ ($h$ and $\chi$ 
Persei, Schmidt-Kaler 1967; NGC~3603, van den Bergh 1978).  Although it 
cannot be excluded, of course, that a few highly luminous clusters might 
be hiding in remote parts of the Galactic disk, it appears unlikely that 
the Milky Way contains any significant number of highly luminous clusters 
similar to those in M51. The brightest clusters in M51 even outshine
the R136 cluster at the centre of the 30 Dor region in the LMC which 
has $M_V = -10.6$ \cite{van78}. 

  Do YMCs represent a separate class of objects, distinct from low-mass 
open clusters, or can YMC populations be reconciled with normal populations
of low-mass clusters, assuming a standard power-law luminosity function (LF)?
Let us compare the YMC population in M51 with the system of open clusters 
in the Milky Way, for the moment ignoring other differences between the 
two galaxies.  van den Bergh \& Lafontaine \shortcite{van84} studied the 
luminosity function of Milky Way 
open clusters and found it to be well represented by the relation
\begin{equation}
  \log \sigma(M_V) = 1.55 + 0.2 M_V
  \label{eq:oclf}
\end{equation}
for clusters fainter than $M_V = -9$, where $\sigma(M_V)$ is the number 
of clusters per kpc$^2$ per magnitude bin. In luminosity units, this becomes
\begin{equation}
  n(L) \propto L^{\alpha}
\end{equation}
where $L$ is the luminosity, $n(L)$ is the number of clusters per
luminosity bin and $\alpha = -1.5$ \cite{elm97}. 
  In the Milky Way, extrapolation of (\ref{eq:oclf}) to higher luminosities 
would yield a total of $\sim 100$ clusters with $M_V = -11$, assuming that 
the Galactic disk has an area of 500 kpc$^2$. This is clearly incompatible 
with what is actually observed and van den Bergh \& Lafontaine 
\shortcite{van84} suggested that the luminosity
function of open clusters in the Milky Way drops below relation 
(\ref{eq:oclf}) somewhere in the magnitude range $-11 < M_V < -8$. 
In fact, there {\it must} be a drop-off in a luminosity function of the
form (\ref{eq:oclf}) at some magnitude level, as the total luminosity of 
a cluster system with a power-law LF diverges for $\alpha > -2$.  However,
it is less obvious that the drop-off will occur at the same magnitude in
all galaxies.

  Young LMC clusters seem to follow a similar power-law LF with 
$\alpha = -1.5 \pm 0.2$, but the LF of young clusters declines more slowly 
at bright magnitudes than in the Milky Way \cite{els85} due to the 
LMC's population of YMCs. So at least in the LMC, YMCs appear to mark
a natural extension of the open cluster population to brighter magnitudes.

  M51 contains 23 and 13 clusters in the two magnitude intervals 
$-11 < M_V < -10$ and $-12 < M_V < -11$, respectively. If the 
$\Delta V = 0.38$ correction is applied, the numbers are $20$ clusters 
with $-11 < M_V < -10$ and $6$ clusters with $-12 < M_V < -11$.
Most of the visible part of the disk of M51 is confined within a diameter 
of about $6^\prime$, corresponding to an area of 170 kpc$^2$.  
Applying Eq.~(\ref{eq:oclf}) 
to the two magnitude intervals above, 60 and 38 clusters would 
be expected -- three times more than what is actually observed.  Of course, the 
extrapolation of Eq.~(\ref{eq:oclf}) to such bright magnitudes is very 
sensitive to the exact value of the exponent $\alpha$.  Changing $\alpha$ 
to $-2$ instead of $-1.5$ and using the Milky Way clusters in the magnitude 
bin $-5 < M_V < -8$ for reference (5.2 clusters kpc$^{-2}$, van den Bergh 
and Lafontaine 1984), the predicted numbers drop to 6 clusters 
with $-11<M_V<-10$ and 2 in the interval $-12 < M_V < -11$, now several
times {\it below} the observed numbers.

  Although very crude, these calculations show that even though M51 is 
apparently much richer in {\it luminous} clusters than the Milky Way, the 
number of low-luminosity clusters in the two galaxies could still be quite 
similar without requiring any particularly peculiar luminosity function.  
The YMCs in M51 could easily be accomodated by simple extrapolation 
of a power-law luminosity function with an exponent somewhere between
$-1.5$ and $-2$, and there is apparently no need to conceive them as
a separate class of objects. 

\subsection{Formation of massive clusters}

  From Fig.~\ref{fig:sfr_tlu}, M51 is indeed quite rich in YMCs compared 
to the ``average'' spiral galaxy.  However, other spiral galaxies with 
similar rich YMC populations are known, and there is no particular evidence 
that the presence of YMCs in M51 is boosted by the interaction with NGC~5195.
If anything, M51 may be slightly cluster-poor for its \ssfr .
None of the other cluster-rich galaxies in the LR2000 sample show obvious 
signs of interaction, and the differences in their SFRs and \tlu\ values 
may be explained simply by different amounts of gas available for star 
formation \cite[LR2000]{ken98}.  M51 itself is also very gas-rich 
(see e.g.~table 1 in Kennicutt 1998b), consistent with its relatively 
high \ssfr\ and \tlu\ values.

   Even higher levels of star forming activity are seen in merger
galaxies like the Antennae \cite{whi99} and NGC~3256 \cite{zep99}, but
also in starbursts that are not directly related to merger events like
e.g. M82 \cite{oco95} and NGC~5253 \cite{gor96}. Such galaxies contain
large numbers of highly luminous young clusters, emitting up to 15\% of 
the blue or UV light \cite{meu95,zep99} and mark an extension of the
\ssfr --\tlu\ relation (LR2000).  At the other extreme of the relation
are galaxies like IC~1613 with exceedingly low (but non-vanishing) 
star formation rates and very few star clusters at all \cite{wyd00}.  

  It thus appears that galaxies form YMCs {\it whenever the SFR is high 
enough}.  This may have important implications for the understanding of 
how massive clusters (aka {\it globular clusters}) formed in the early 
Universe.  At earlier epochs, the general level of star formating activity 
was presumably higher because of the larger amounts of gas available, and 
globular clusters may have formed quite naturally. The highest levels of 
star forming activity may have existed at the centres of rich galaxy 
clusters, giving rise to the rich populations of globular clusters seen 
around many cD galaxies.

  The key to understanding massive cluster formation may lie in the
properties of the interstellar medium in their parent galaxies.
In the Milky Way, star clusters form in the cores of highly fragmented 
giant molecular clouds (GMCs) \cite{lad97}, but at a very low 
efficiency averaged over the entire GMC. The largest GMCs in the Milky 
Way have masses of about $5\times10^6 \msun$ with a sharp cut-off above 
this limit \cite{mck99}. It is tempting to speculate that this upper 
cut-off may be related to the upper limit of the open cluster luminosity 
function discussed in the previous section.  If massive clusters form with 
similar low efficiencies, their parent clouds must be much larger than 
Milky Way GMCs and it has been suggested that globular clusters formed from
{\it supergiant molecular clouds} (SGMCs) with masses of $10^8 - 10^9 \msun$
\cite{har94}.  In M51 and NGC~5236, high-resolution CO studies have in
fact revealed ``Giant Molecular Associations'' (GMAs) with masses 
of $10^7 - 10^8 \, \msun$ \cite{vog88,ran90,ran99}, which might be 
identifiable as the birthsites of YMCs. Recently, 
Wilson et al.~\shortcite{wil00} reported even more massive GMAs with
masses of $3 - 6\times10^8\, \msun$ in the YMC--rich ``Antennae'' galaxies.
Such GMAs may assemble more
easily in galaxies with high gas densities and, consequently, high star
formation rates \cite{ken98}. The connection between SFR and YMC 
richness could then be understood as resulting from an underlying dependence
of both on the gas density.

\section{Conclusions}
\label{sec:conc}

  Young Massive Star Clusters (YMCs) have been identified in the nearby
face-on Sbc-type spiral M51. The richness of the YMC population in
M51, as measured by the specific $U$-band luminosity \tlu\, is
comparable to other cluster-rich spiral galaxies, compatible with the high
star formation rate in M51 and the \ssfr\ -- \tlu\ relation \cite{lar00}.
The high level of star formation activity may in this particular case be 
partly stimulated by interaction with the nearby companion NGC~5195, but 
interactions do not in general seem to be a necessary requirement for YMC 
formation.

  Within order-of magnitude estimates, the number of YMCs in M51 is 
compatible with extrapolation of a power-law luminosity function with 
exponent $\alpha \sim -2$ from a population of low-mass open clusters 
similar to that in the Milky Way.  Thus, YMCs may simply represent a 
continuation of the normal open cluster luminosity function, extending 
to different upper limits in different galaxies. Therefore YMCs plausibly
form in much the same way as lower-mass clusters and it appears likely that 
these same basic mechanisms also applied to the formation of globular
clusters in the early Universe.

\section{Acknowledgments}

  This work was supported by National Science Foundation grant number
AST9900732 and Faculty Research funds from the University of California,
Santa Cruz.

\begin{table}
\caption{\label{tab:clusters} 
   Data for clusters in M51. $M_V$, \ub\ and \bv\ have been
corrected for a foreground extinction of $A_B = 0.150$ mag. Clusters
marked with an asterisk ($*$) were included on the HST F439W and F555W
pointings.
}
\begin{tabular}{rrrrrr}
  ID & RA(2000.0) & DEC(2000.0) & $M_V$ & \ub & \bv \\ \hline
  27 & 13:29:35.09 & 47:12:12.5 & $-9.32$ & $-0.36$ & $0.38$ \\
  31 & 13:29:35.70 & 47:10:44.4 & $-9.51$ & $-0.38$ & $0.30$ \\
  37 & 13:29:36.54 & 47:09:13.0 & $-10.51$ & $-0.84$ & $0.30$ \\
  41 & 13:29:36.80 & 47:11:19.3 & $-10.24$ & $-0.48$ & $0.29$ \\
  46 & 13:29:37.14 & 47:10:04.2 & $-10.09$ & $-0.56$ & $0.39$ \\
  61 & 13:29:38.31 & 47:11:05.1 & $-10.42$ & $-0.65$ & $0.19$ \\
  84 & 13:29:39.78 & 47:10:32.9 & $-10.62$ & $-0.69$ & $0.17$ \\
  92 & 13:29:40.04 & 47:09:24.0 & $-9.58$ & $-0.35$ & $0.28$ \\
  94 & 13:29:40.16 & 47:10:43.8 & $-10.37$ & $-0.65$ & $0.14$ \\
 101 & 13:29:40.57 & 47:11:51.2 & $-10.82$ & $-0.34$ & $0.28$ \\
 111 & 13:29:41.27 & 47:13:25.8 & $-9.46$ & $-0.17$ & $0.37$ \\
 122 & 13:29:41.85 & 47:11:42.6 & $-10.43$ & $-0.42$ & $0.20$ \\
 130 & 13:29:42.40 & 47:13:19.0 & $-9.58$ & $-0.46$ & $0.18$ \\
 162 & 13:29:43.64 & 47:10:48.5 & $-9.66$ & $-0.74$ & $0.17$ \\
 172 & 13:29:43.90 & 47:09:55.6 & $-10.92$ & $-0.92$ & $0.32$ \\
 176 & 13:29:43.99 & 47:09:32.8 & $-9.68$ & $-0.97$ & $0.37$ \\
 180 & 13:29:44.13 & 47:10:23.7 & $-12.46$ & $-1.11$ & $0.07$ \\
 203 & 13:29:44.94 & 47:09:59.4 & $-12.82$ & $-1.08$ & $0.08$ \\
 263 & 13:29:46.52 & 47:12:33.5 & $-11.45$ & $-0.97$ & $0.06$ \\
 320 & 13:29:48.09 & 47:13:32.9 & $-8.80$ & $-0.37$ & $-0.02$ \\
 330 & 13:29:48.47 & 47:13:19.6 & $-9.72$ & $-0.67$ & $0.32$ \\
 335 & 13:29:48.73 & 47:13:34.7 & $-9.61$ & $-0.68$ & $0.05$ \\
 368 & 13:29:49.95 & 47:08:34.3 & $-9.00$ & $-0.34$ & $0.32$ \\
 395 & 13:29:50.66 & 47:10:07.1 & $-11.12$ & $-0.88$ & $0.11$ \\
 $*$403 & 13:29:50.83 & 47:10:41.4 & $-10.13$ & $-0.84$ & $0.34$ \\
 409 & 13:29:51.15 & 47:09:18.0 & $-9.87$ & $-0.92$ & $0.11$ \\
 416 & 13:29:51.47 & 47:10:10.1 & $-9.57$ & $-0.79$ & $0.11$ \\
 420 & 13:29:51.67 & 47:09:09.2 & $-10.46$ & $-0.89$ & $0.19$ \\
 422 & 13:29:51.69 & 47:08:52.9 & $-9.69$ & $-0.95$ & $0.15$ \\
 459 & 13:29:52.77 & 47:09:12.7 & $-9.84$ & $-0.92$ & $0.34$ \\
 $*$477 & 13:29:53.22 & 47:12:39.7 & $-11.30$ & $-0.77$ & $0.21$ \\
 $*$479 & 13:29:53.30 & 47:10:42.9 & $-11.60$ & $-0.90$ & $0.10$ \\
 495 & 13:29:53.83 & 47:12:57.7 & $-9.94$ & $-0.21$ & $0.31$ \\
 $*$534 & 13:29:54.88 & 47:12:50.6 & $-10.23$ & $-0.79$ & $0.18$ \\
 559 & 13:29:55.55 & 47:14:02.5 & $-11.58$ & $-1.01$ & $0.01$ \\
 $*$589 & 13:29:56.30 & 47:12:43.2 & $-10.31$ & $-0.74$ & $0.07$ \\
 $*$597 & 13:29:56.55 & 47:10:47.6 & $-9.81$ & $-0.77$ & $0.18$ \\
 $*$627 & 13:29:57.05 & 47:11:31.7 & $-9.58$ & $-0.99$ & $0.24$ \\
 649 & 13:29:57.64 & 47:09:32.5 & $-9.21$ & $-0.12$ & $0.26$ \\
 684 & 13:29:58.48 & 47:13:49.4 & $-10.63$ & $-1.07$ & $-0.09$ \\
 $*$693 & 13:29:58.65 & 47:12:58.3 & $-10.29$ & $-0.59$ & $0.15$ \\
 $*$713 & 13:29:58.96 & 47:11:04.7 & $-11.38$ & $-0.98$ & $0.07$ \\
 747 & 13:29:59.70 & 47:13:40.6 & $-11.29$ & $-0.94$ & $0.04$ \\
 750 & 13:29:59.70 & 47:13:59.2 & $-12.54$ & $-0.96$ & $0.04$ \\
 759 & 13:29:59.90 & 47:09:36.3 & $-9.36$ & $-0.22$ & $0.40$ \\
 803 & 13:30:00.60 & 47:13:27.0 & $-10.93$ & $-0.79$ & $0.12$ \\
 832 & 13:30:00.97 & 47:09:29.5 & $-10.86$ & $-1.37$ & $0.05$ \\
 839 & 13:30:01.12 & 47:13:45.3 & $-11.09$ & $-0.81$ & $0.02$ \\
 842 & 13:30:01.27 & 47:16:53.9 & $-10.62$ & $-0.72$ & $0.20$ \\
 848 & 13:30:01.32 & 47:12:51.5 & $-11.52$ & $-0.97$ & $0.03$ \\
 852 & 13:30:01.35 & 47:09:12.6 & $-8.53$ & $-0.01$ & $0.23$ \\
 872 & 13:30:01.84 & 47:15:34.4 & $-10.19$ & $-0.28$ & $0.36$ \\
 883 & 13:30:02.11 & 47:13:26.1 & $-10.78$ & $-0.67$ & $0.08$ \\
 890 & 13:30:02.25 & 47:12:33.1 & $-10.31$ & $-0.51$ & $0.12$ \\
 899 & 13:30:02.43 & 47:09:49.6 & $-12.18$ & $-1.12$ & $0.20$ \\
 911 & 13:30:02.78 & 47:09:57.3 & $-10.99$ & $-1.18$ & $0.16$ \\
 $*$912 & 13:30:02.82 & 47:11:30.1 & $-9.93$ & $-0.58$ & $0.32$ \\
 924 & 13:30:03.18 & 47:12:45.5 & $-11.01$ & $-1.00$ & $-0.01$ \\
 936 & 13:30:03.81 & 47:12:00.9 & $-9.60$ & $-0.96$ & $0.04$ \\
 947 & 13:30:03.97 & 47:10:15.3 & $-11.18$ & $-0.87$ & $0.13$ \\
\end{tabular}
\end{table}

\begin{table}
Table~\ref{tab:clusters} (continued) \\
\begin{tabular}{rrrrrr}
  ID & RA(2000.0) & DEC(2000.0) & $M_V$ & \ub & \bv \\ \hline
 984 & 13:30:05.03 & 47:12:33.1 & $-11.12$ & $-0.88$ & $0.08$ \\
 987 & 13:30:05.04 & 47:10:18.6 & $-9.52$ & $-1.03$ & $0.07$ \\
 993 & 13:30:05.19 & 47:10:50.5 & $-8.85$ & $-0.22$ & $0.27$ \\
1013 & 13:30:05.70 & 47:09:49.9 & $-9.90$ & $-0.32$ & $0.27$ \\
1019 & 13:30:05.92 & 47:11:40.5 & $-9.82$ & $-0.62$ & $0.35$ \\
1031 & 13:30:06.55 & 47:11:01.1 & $-9.51$ & $-0.75$ & $0.21$ \\
1033 & 13:30:06.67 & 47:14:20.7 & $-11.57$ & $-0.86$ & $0.08$ \\
1049 & 13:30:07.40 & 47:11:26.5 & $-10.12$ & $-1.01$ & $0.03$ \\
1066 & 13:30:08.47 & 47:13:59.7 & $-9.34$ & $-0.22$ & $0.36$ \\ \hline
\end{tabular}
\end{table}

\end{document}